\documentclass[epj]{svjour}
\usepackage{graphicx}
\usepackage{dcolumn}
\usepackage{bm}
\begin{document}
\title{Self-consistent calculations within the Green's function method including particle-phonon coupling 
       and the single-particle continuum}
\titlerunning{Self-consistent calculations}

\author{N. Lyutorovich\inst{1,2} \and J. Speth\inst{1,3} \mail{j.speth@fz-juelich.de} \and 
        A. Avdeenkov\inst{1,4} \and F. Gr\"ummer\inst{1} \and S. Kamerdzhiev\inst{4} \and 
        S. Krewald\inst{1} \and V.I. Tselyaev\inst{2}}
\authorrunning{N Lyutorovich et al.}

\institute{Institut f\"ur Kernphysik, Forschungszentrum J\"ulich, 52425 J\"ulich, Germany \and
           Institute of Physics S.Petersburg University, Russia \and
           Institute of Nuclear Physics, PAN, PL-31-342 Cracow, Poland \and
           Institute of Physics and Power Engineering, 249020 Obninsk, Russia}

\date{Received: date / Revised version: date}

\abstract{
The Green's function method in the \emph{Quasiparticle Time Blocking Approximation} is applied to nuclear excitations in $^{132}$Sn and $^{208}$Pb. 
The calculations are performed self-consistently using a Skyrme interaction. 
The method combines the conventional RPA with an exact single-particle continuum treatment and considers in a consistent way the particle-phonon coupling. 
We reproduce not only the experimental values of low- and high-lying collective states but we also obtain fair agreement with the data of non-collective low-lying states that are strongly influenced by the particle-phonon coupling.
}

\maketitle

\section{Introduction}
Theoretical investigations of unstable neutron rich nuclei require nuclear mean fields in experimentally unexplored mass regions. 
Therefore we have generalized an approach formulated in the many-body Green's function formalism which includes single-particle degrees of freedom and in a consistent way collective nuclear excitations \cite{kst93,rev04,Vic} by incorporating self-consistent nuclear mean fields derived from Skyrme interactions.
The exact treatment of the continuum is the main difference of the present approach compared to a previous paper \cite{Avd}. 
The continuum is first of all important for the distribution of the high-lying multipole strength but it also allows a clear definition of the size of the configuration space. 
We demonstrate these effects by comparing the present results obtained with the exact continuum treatment and previous results, where the continuum was discretized. 
An equally important point concerns the energies of the low-lying non-collective states. 
Here we investigate the influence of the phonons on such excitations. 
Whereas the low-lying collective states are influenced by the single-particle spectrum and the residual particle-hole (ph) interaction, the non-collective states obtained in RPA calculations are only little affected by the interaction and are very close to the unperturbed ph-energies.
But these states are influenced by the particle-phonon coupling.
Calculations \cite{Ring74} performed with the Migdal approach \cite{Migdal67}, where experimental single-particle energies are used, gave good agreement with the data for collective and non-collective states as the latter ones are close to the unperturbed experimental ph-energies. 
This is different in self-consistent calculations, where the single-particle spectrum depends on the effective mass m$^*$.
The various Skyrme parametrizations, which all reproduce the ground-state properties very well, differ considerably in the effective mass. 
Many self-consistent nuclear structure calculations are performed with parameter sets which correspond to effective masses in the vicinity of m$^*$/m $\approx$ 0.7-0.8. 
It is obvious that in those cases the collective states may be reproduced correctly but the non-collective ones will differ appreciably from the experimental value. 
All this will be discussed in sections three and four. 
In the next section we give a short introduction into the present approach and compare it with existing models. 
In the last section we summarize the results and draw some conclusions. 

\section{Method}
We use the \emph{Quasiparticle Time Blocking Approximation} (QTBA) formulated in \cite{Vic}, see also the preceding work \cite{kst93}. Some details of the self-consistent application are given in \cite{Avd}. 
Here we only summarize the crucial equations. 
The conventional 1p1h \emph{Random Phase Approximation} (RPA) written in the configuration space of the single-particle wave functions ${\varphi_\nu}$ has the form:
\begin{eqnarray}\label{eq:1}
\left( \epsilon_{\nu_1}-\epsilon_{\nu_2}- \Omega\right)\chi^{m}_{\nu_1 \nu_2} = \\ \nonumber
\left(n_{\nu_1}-n_{\nu_2}\right)\sum_{\nu_3 \nu_4}F^{ph}_{\nu_1 \nu_4 \nu_2 \nu_3} \;\chi^{m}_{\nu_3 \nu_4}
\,.
\end{eqnarray}
The $\epsilon_{\nu}$ are the single-particle energies, $F^{ph}$ is the ph-interaction, $n_\nu$ are the occupation numbers: $0$ and $1$ for particles and holes, respectively. 
In the self-consistent approach all these input data follow from the mean field solution. 
From equation (\ref{eq:1}) one obtains the excitation energy $\Omega$ of an even-even nucleus and the corresponding ph-transition matrix elements  $\chi^{m}_{\nu_1 \nu_2}$. 

For the numerical applications it is more convenient to solve the equation in the $\bf{r}$-space because 
this allows the treatment of the continuum in the most efficient way. 
Instead of the homogeneous integral equation (\ref{eq:1}) one solves an inhomogeneous equation of the form:
\begin{eqnarray}\label{eq:2}
\rho(\bm{r},\Omega) = -\int d^{3}\bm{r}' A(\bm{r},\bm{r}',\Omega)
Q^{\rm eff}(\bm{r}',\Omega) \\ \nonumber -\int
d^3\bm{r}'d^3\bm{r}''A(\bm{r},\bm{r}',\Omega)
F^{ph}(\bm{r}',\bm{r}'') \rho(\bm{r}'',\Omega)\,.
\end{eqnarray}
where $Q^{\rm eff}(\bm{r}',\Omega)$ is an external field and  $A(\bm{r},\bm{r}',\Omega)$ is the ph-propagator in the $\bf{r}$-space. 
The poles of this equation are the excitation energies of an even-even nucleus and $\rho(\bm{r},\Omega)$ at a given pole is the corresponding transition density.

\begin{figure}[htbp]
\begin{center}
\includegraphics[bb=44 364 405 551,width=8cm]{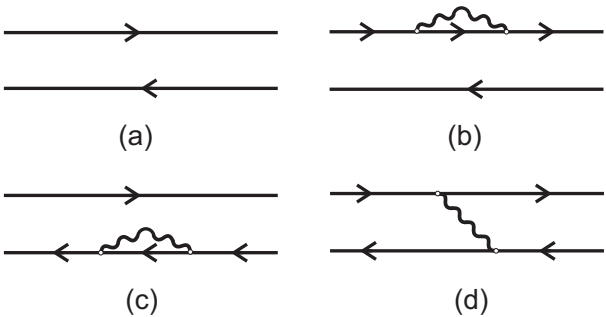}
\end{center}
\caption{\label{fig:1}Graph (a) denotes the uncorrelated ph-propagator of the RPA. 
Corrections due to phonons are indicated in (b-d). 
The graphs (b) and (c) are corrections to the propagator and (d) is a contribution to the ph-interaction. 
The wavy lines and the solid lines denote the phonons and the single-particle propagators, respectively.}
\end{figure}

The inclusion of phonons gives rise to a modification of the single-particle energies as indicated in Fig. \ref{fig:1}(b) for particles and Fig. \ref{fig:1}(c) for holes. 
It also gives rise to a modification of the ph-interaction. 
This is shown in Fig. \ref{fig:1}(d). 
The QTBA version which we used here considers some more corrections. 
A detailed derivation is given in Refs. \cite{Vic,LVic}. 
A further advantage of the $\bf{r}$-space is that the structure of Eq.(\ref{eq:2}) is not changed if the phonons are included, only the propagator $A$ and the interaction $F^{ph}$ are modified.

It is important to point out that no new parameters appear if one introduces phonons. 
The modified single-particle energies $\widetilde{\epsilon}_{\nu}$ can be calculated from the formula:

\begin{equation}\label{eq:3}
\widetilde{\epsilon}_{\nu_1} =\epsilon_{\nu_1} +\sum_{\nu_2 ;i}
 \frac{\left| \gamma^{\nu_2 ;i}_{\nu_1}\right|^2}
{\widetilde{\epsilon}_{\nu_1}   -\Omega_i  -\epsilon_{\nu_2}},
\end{equation}
where the second term on the right side is graphically represented by the upper part of graph (b) in Fig. \ref{fig:1}.
The vertex $\gamma_{\nu} ^{\mu;i}$, which couples the quasi-particle state $\nu$ to the core excited configuration $\mu{\otimes}i$, is given by:
\begin{equation}\label{eq:4}
\gamma_{\nu} ^{\mu;i} = \sum_{\alpha,\beta} F^{ph}_{\nu \alpha; \mu \beta} \chi^{i}_{\alpha\beta},
\end{equation}
where $\chi^i$ is the RPA wave function (Eq. (\ref{eq:1})) of the phonon considered.
The corresponding energy-dependent correction to the ph-interaction represented by graph(d) in Fig. \ref{fig:1} has the form
\begin{equation}\label{eq:5}
F^{ph,phonon}_{\alpha \mu, \beta \nu}(\Omega,\epsilon,\epsilon') \nonumber
= \sum_{i}\frac{(\gamma^{\mu;i}_{\alpha})^* \gamma^{\nu;i}_{\beta}}{ \epsilon - \epsilon'+  (\Omega_i-i\delta)}.
\end{equation}
All input data are obtain from the mean field and the conventional 1p1h RPA, respectively. 
Other corrections can be calculated in a similar way.

\section{Numerical details}
We calculate the mean field within the HFB approach with the code HFBRAD \cite{bennaceur} and use in all applications the SLy4 \cite{SLy4} parametrization of the Skyrme ansatz. 
The excited states are calculated with a computer code which is based on the the QTBA version \cite{Vic,LVic} of the extended theory of finite fermi systems (ETFFS) \cite{kst93,rev04}.
Compared to our previous calculations \cite{Avd} we have modified the code in such a way that the single-particle continuum is treated exactly. 
In self-consistent calculations the ph-interaction is given by the second derivative of the energy functional. We employ the same approximations as in Ref. \cite{Avd}.  
We apply the same procedure to the theoretical distributions \cite{kst93}. 
The theoretical mean energies E$_{0}$, resonance widths ${\Gamma}$ and maximum values of the cross section $\sigma_{0}$ are extracted under the condition that the first three energy-weighted moments of the Lorentzian and the theoretical distribution of the total photoabsorption cross section should coincide.

\section{Results}
\subsection{Low-lying collective and non-collective states}
When one solves the RPA equation one obtains as many solutions as one has (discrete) ph-components. 
From the schematic model by Brown and Bolsterli \cite{BB} one knows that there is one collective state for each multipolarity and isospin which is strongly shifted in energy, whereas the remaining states are little influenced by the interaction. 
As the isoscalar interaction is attractive the first excited state of each multipolarity (in RPA and QTBA) is shifted to lower energies, whereas the other states remain close to their unperturbed ph-energies. 
Beyond this shift due to the ph-interaction, there is an additional energy shift in QTBA which is due to the phonon coupling. This affects all states.
As discussed in \cite{Avd} the phonons compress the single-particle spectrum and therefore the unperturbed energies in the propagator of the QTBA are compressed compared to the HF energies. 
This effect and its consequences are the main issue of this chapter. 

\begin{table}[!hb]
\caption{\label{tab:t1}The four lowest $3^{-}$ states in $^{208}$Pb. 
Here we compare the results with and without phonons for the bare ph-energies and the solutions of RPA and QTBA. 
In both cases the single-particle continuum is treated exactly.} 
\begin{center}
\begin{tabular}[b]{|c|c|c|c|c|}
\hline
  \ \ \ HF \ \ \ & HF+Ph &\ \  RPA \ \   & QTBA  &  \ \ \ Exp. \ \ \   \\
\hline
 5.76      & 4.30     &3.25 & 2.43 &    2.61 \\
 5.93      & 4.93     &5.96 & 4.53 &    4.05 \\
 6.03      & 4.95    &6.18& 4.96 &      4.25   \\
 6.68      & 5.17    &6.65 & 5.13 &     4.70    \\ 
\hline
\end{tabular}
\end{center}
\end{table}

In Tables \ref{tab:t1}-\ref{tab:t4} we show in the first column (HF) the unperturbed ph-energies calculated with the SLy4 parametrization. 
In the next column (HF+Ph) the unperturbed energies of the QTBA propagators are shown that are compressed due to the phonon coupling. 
In the third and fourth column we present the RPA and QTBA results, respectively. 

\begin{table}[!hb]
\caption{\label{tab:t2}Same as in Table \ref{tab:t1}, but for the four lowest $5^{-}$ states in $^{208}$Pb.} 
\begin{center}
\begin{tabular}[b]{|c|c|c|c|c|}
\hline
   \ \ \ HF \ \ \ & HF+Ph &\ \  RPA \ \   & QTBA  &  \ \ \ Exp. \ \ \   \\
\hline
4.93       & 3.97     &4.13 & 3.30 &    3.20 \\
5.01       & 4.03     &5.06 & 4.07 &    3.71 \\
5.76       & 4.92     &5.48&  4.52 &    3.96   \\
5.93       & 4.9     &5.92 & 4.99 &    4.13   \\ 
\hline
\end{tabular}
\end{center}
\end{table}

We first discuss the levels with the multipolarity $3^-$ in Table \ref{tab:t1}. 
The first excited state is the most collective one in $^{208}$Pb and therefore shows the strongest shift due to the ph-interaction. 
The other levels are only little affected by the interaction. 
This holds for RPA and QTBA and can been seen if one compares column I and column III (for RPA) and column II and column IV (for QTBA), respectively. 
One may argue that the agreement between theory and experiment for the lowest state calculated in RPA is not too bad. 
But it is clear that the energies of the other (RPA)-states strongly deviate from the experiment. 
This situation is typical for all effective Lagrangians with an effective mass less then one. 
In the present case, where one uses m$^*$/m = 0.70, the compression due to the phonons is of crucial importance for a quantitative description of collective and non-collective states. 
The compression of the spectra for the four lowest $3^-$ configurations can be deduced from column I and column II. 
One realizes that the phonons compress the spectra by more then $1$MeV which strongly improves the agreement between our QTBA results and the data for the collective as well as the non-collective state. 
This is also true for the other multipolarities which are shown in table \ref{tab:t2} - table \ref{tab:t4}. 

\begin{table}[!hb]
\caption{\label{tab:t3}Same as in Table \ref{tab:t1}, but for the four lowest $2^{+}$ states in $^{208}$Pb.} 
\begin{center}
\begin{tabular}[b]{|c|c|c|c|c|}
\hline
   \ \ \ HF \ \ \ & HF+Ph &\ \  RPA \ \   & QTBA  &  \ \ \ Exp. \ \ \   \\
\hline
5.91   & 5.052    &5.33 & 4.39 &    4.08 \\
6.43   & 5.29     &6.20 & 5.26 &    4.93 \\
6.83   & 5.51    &6.42&  5.96 &    5.56   \\
7.98   & 7.360   &7.12 & 6.43 &    5.72    \\ 
\hline
\end{tabular}
\end{center}
\end{table}

\begin{table}[!hb]
\caption{\label{tab:t4}Same as in Table \ref{tab:t1}, but for the four lowest $4^{+}$ states in $^{208}$Pb.} 
\begin{center}
\begin{tabular}[b]{|c|c|c|c|c|}
\hline
   \ \ \ HF \ \ \ & HF+Ph &\ \  RPA \ \   & QTBA  &  \ \ \ Exp. \ \ \   \\
\hline
5.91       & 5.05     &5.59 & 4.61 &    4.32   \\
6.43       & 5.39     &6.17 & 5.17 &    4.91   \\
6.83       & 5.63     &6.77 & 5.52 &    5.22   \\
7.98       & 7.36     &8.05 & 7.18 &    5.56    \\ 
\hline
\end{tabular}
\end{center}
\end{table}

\begin{table}[!hb]
\caption{\label{tab:t5}Low-lying collective states in $^{132}$Sn. 
Here we compare the results with and without phonons and with the correct single-particle continuum and discretized continuum states with the data. 
We show the first excited state of each multipolarity.}
\begin{center}
\begin{tabular}[b]{|c|c|c|c|c|c|}
\hline
  & \multicolumn{2}{c|}{full continuum} & \multicolumn{2}{c|}{disc.basis} & \\
\hline
  & \ \  RPA \ \   & QTBA  & \ \  RPA \ \ & QTBA & \ \ \ Exp. \ \ \   \\
\hline
2$^{+}$ & 4.46 & 3.70 & 4.46  & 2.10 & 4.04 \\
3$^{-}$ & 5.19 & 4.60 & 5.08  & 3.00 & 4.35 \\
4$^{+}$ & 5.00 & 4.20 &  --   & 2.85   & 4.42 \\ 
\hline
\end{tabular}
\end{center}
\end{table}

\begin{table}[!hb]
\caption{\label{tab:t6}Giant dipole resonance in $^{132}$Sn. 
Here we compare the results with and without phonons and with the correct single-particle continuum and discretized continuum states with the data \cite{GSI}. 
In the case of QTBA* the isovector force has be increased by 10\%.}
\begin{center}
\begin{tabular}[b]{|c|c|c|c|c|}
\hline
  & \multicolumn{3}{c|}{full continuum} & \\
\hline
  & \ \  RPA \ \   & QTBA  & QTBA* & \ Exp. \ \\
\hline
E(MeV) & 14.3 & 13.8 & 16.1& 16.1(7) \\
$\Gamma$(MeV) & 3.4 & 4.0 & 4.5& 4.7(2.1) \\
\hline
  & \multicolumn{3}{c|}{discrete basis} & \\
\hline
  & \ \  RPA \ \   & QTBA  & & \\
\hline
E(MeV) & 14.3  & 12.5 & & 16.1(7) \\
$\Gamma$(MeV) & 3.8  & 5.2 & & 4.7(2.1) \\
\hline
\end{tabular}
\end{center}
\end{table}

\begin{table}[!hb]
\caption{\label{tab:t7}Giant dipole resonance in $^{208}$Pb. 
Here we compare the results with and without phonons with the data. 
In both cases the single-particle continuum is treated exactly. 
In the case of QTBA* the isovector force was increased by 6\%.} 
\begin{center}
\begin{tabular}[b]{|c|c|c|c|c|}
\hline
  & \ \  RPA \ \   & QTBA  & \  \ QTBA*& \ \ \ Exp. \ \ \   \\
\hline
E(MeV)& 12.4 & 11.8 & 13.4 &  13.4 \\
$\Gamma$ (MeV)& 2.7 & 5.3 & 5.3&  4.1 \\
$\sigma$ (mb)&986& 552 & 544&    640  \\                                                                                      \hline
\end{tabular}
\end{center}                                                                                                   \end{table}

\begin{figure}[htbp]
\vskip 4mm
\begin{center}
\includegraphics[width=8cm]{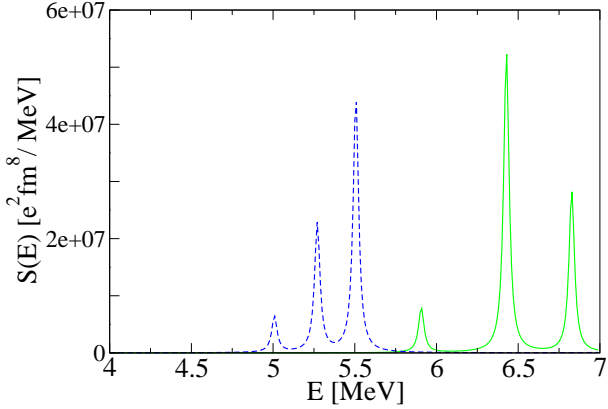}
\end{center}
\caption{\label{fig:2} The three lowest ph-energies for the $2^+$ ( and $4^+$) multipolarity in $^{208}$Pb, in the HF case (full line) and the HF+Ph case (dashed).}
\end{figure}

The effect of the phonons is also demonstrated in Fig. \ref{fig:2}. 
Here the three lowest ph-energies for the $2^+$ (which are the same as for the $4^+$) are shown. 
The full line indicates the HF and the dashed ones the HF+Phonons results. 
One notices a shift about 1 MeV due to the phonon coupling. 
The theoretical ph-energies may be compared with the corresponding three lowest experimental energies at $5.06$ MeV, $5.57$ MeV and $5.84$ MeV.

\begin{figure}[htbp]
\vskip 4mm
\begin{center}
\includegraphics[width=8cm]{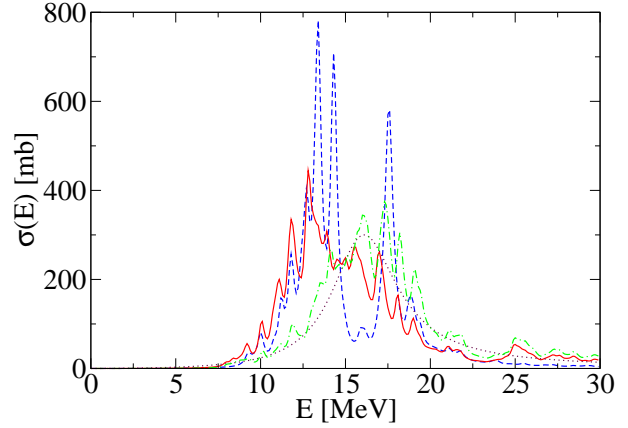}
\end{center}
\caption{\label{fig:3} E1 photoabsorption cross section in $^{132}$Sn. 
The dashed line represents the continuum RPA and the full line the QTBA result. 
The dashed-dotted line indicates the QTBA result where the isovector force was increased by 10\%. 
Here we obtain $E = 16.1$ MeV and $\Gamma = 4.5$ MeV. 
The dotted line is the experimental distribution \cite{GSI}.}
\end{figure}

In \cite{Avd} we published results for low-lying collective states in $^{132}$Sn obtained within the same theoretical framework. 
The same Skyrme parametrization was used but the numerical calculations were done in a discrete configuration space. 
This has consequences for the overlap of the ph wave functions and the renormalization of the interaction (the spurious state has to be at zero energy). 
With the present choice the (compressed) uncorrelated ph excitations are comparable with the experimental values.
In Table \ref{tab:t5} we compare the previous RPA and QTBA results with the present results where the continuum was included in an exact way. 
The present theoretical values agree very nicely with the data, whereas the previous results were much too low.
The comparison shows that there is an arbitrariness in the size of the configuration space if one uses a discretization method and it also demonstrates that one has to chose the number of phonons in such a way that the compressed single-particle spectrum comes close to the experimental one. 
If the continuum is treated exactly the arbitrariness in the size of the configuration space does no longer exist.
Here the SLy4 parametrization seems to be best suited for the application of the QTBA under the present conditions.

\begin{figure}[htbp]
\vskip 4mm
\begin{center}
\includegraphics[width=8cm]{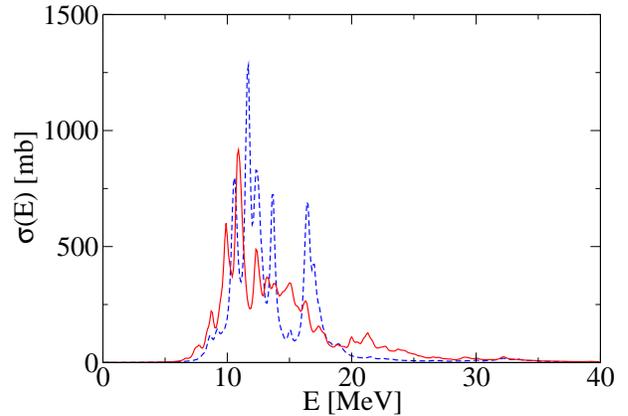}
\end{center}
\caption{\label{fig:4}E1 photoabsorption cross section in $^{208}$Pb. 
The dashed line represents the continuum RPA and the full line the QTBA result.}
\end{figure}

\begin{figure}[htbp]
\vskip 4mm
\begin{center}
\includegraphics[width=8cm]{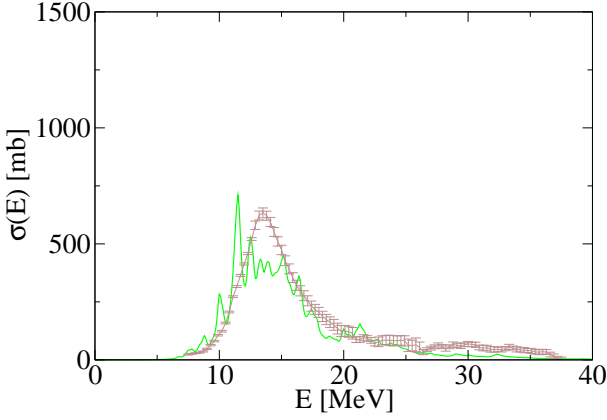}
\end{center}
\caption{\label{fig:5} E1 photoabsorption cross section in $^{208}$Pb. 
The full light line indicates the QTBA result where the isovector force was increased by $6\%$. 
Here we obtain $E = 13.4$ MeV and $\Gamma = 5.3$ MeV. 
The dashed dotted line is the experimental distribution \cite{Fultz69}.}
\end{figure}

\begin{figure}[htbp]
\vskip 4mm
\begin{center}
\includegraphics[width=8cm]{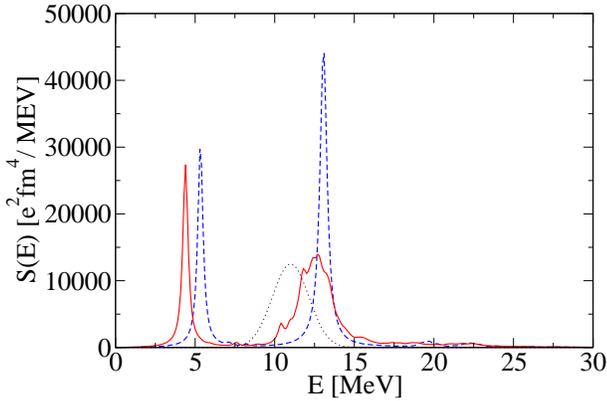}
\end{center}
\caption{\label{fig:6} Isoscalar quadrupole strength in $^{208}$Pb. 
The dashed line represents the continuum RPA and the full line the QTBA result. 
The data are indicated by the dotted line.}
\end{figure}

\begin{figure}[htbp]
\vskip 4mm
\begin{center}
\includegraphics[width=8cm]{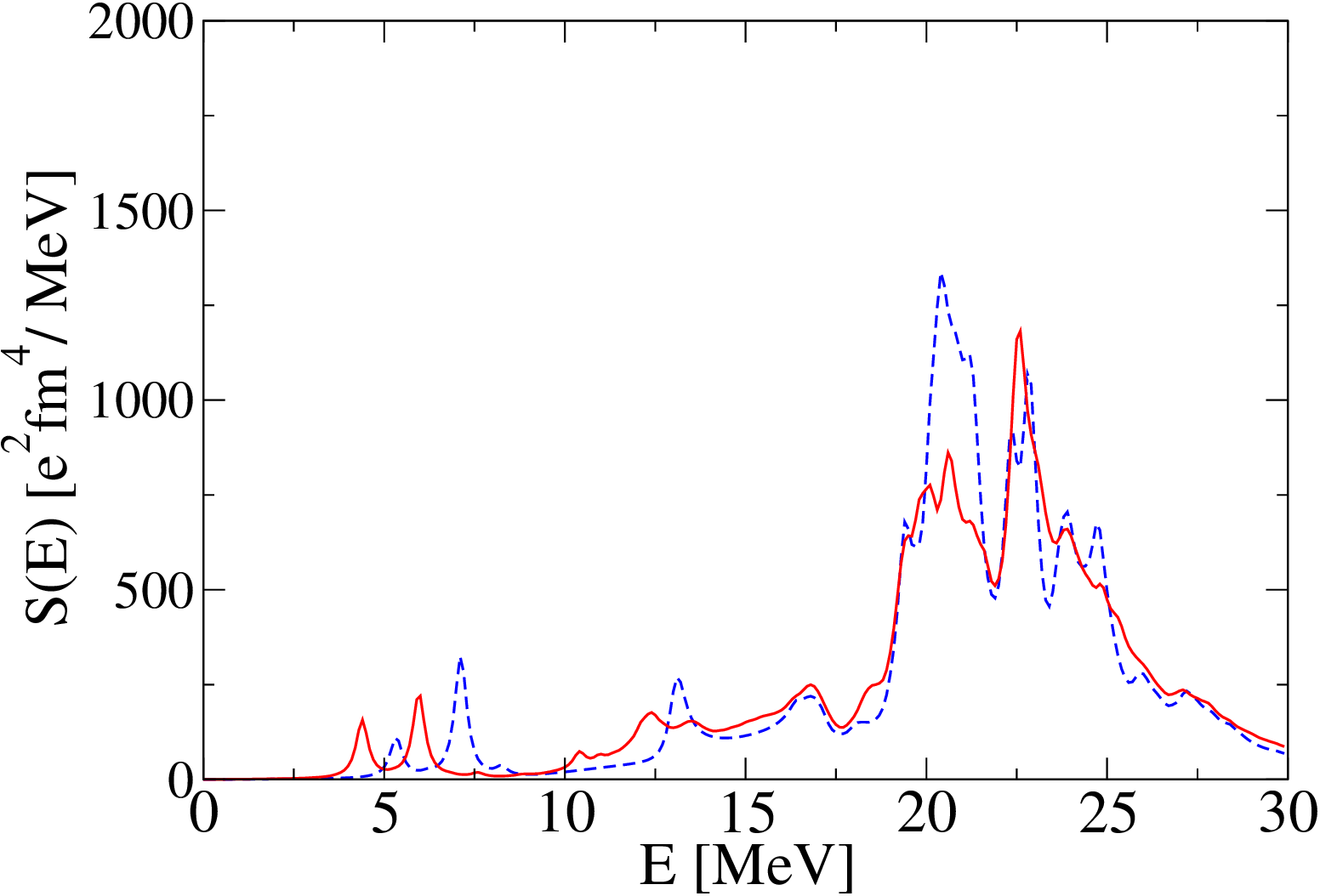}
\end{center}
\caption{\label{fig:7} Isovector quadrupole strength in $^{208}$Pb. 
The dashed line represents the continuum RPA and the full line the QTBA result.}
\end{figure}

\begin{table}[!hb]
\caption{\label{tab:t8}Giant isoscalar and isovector quadrupole resonance in $^{208}$Pb. 
The results with and without phonons are compared  with the data. 
In both cases the single-particle continuum is treated exactly. 
The the isoscalar results (0) are from Ref. \cite{YOU81} and the isovector resonance (1) was measured in $^{209}$Bi \cite{HAK90}.}
\begin{center}
\begin{tabular}[b]{|c|c|c|c|}
\hline
  & \multicolumn{2}{c|}{isoscalar} & \\
\hline
  & \ \ RPA \ \ & QTBA & Exp.(0) \\
\hline
E(MeV) & 13.04 & 12.63 & 11.0(0.2) \\
$\Gamma$(MeV) & 2.38 & 3.30& 3.3(0.3) \\
\hline
  & \multicolumn{2}{c|}{isovector} & \\
\hline
  & \ \ RPA \ \ & QTBA & Exp.(1) \\
\hline
E(MeV) & 22.2  & 22.4 & 22.5(1.0) \\
$\Gamma$(MeV) & 4.00 & 4.78 & 6.0 (3.0) \\
\hline
\end{tabular}
\end{center}
\end{table}

\subsection{Giant Resonances - the high-lying collective states}
\begin{figure}[htbp]
\vskip 4mm
\begin{center}
\includegraphics[width=8cm]{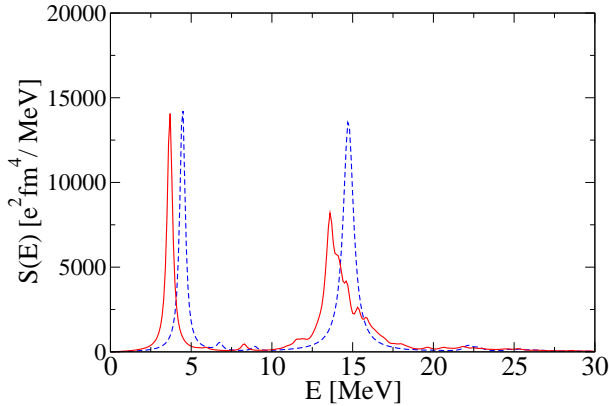}
\end{center}
\caption{\label{fig:8} Isoscalar quadrupole strength in $^{132}$Sn. 
The dashed line represents the continuum RPA and the full line the QTBA result.}
\end{figure}

\begin{figure}[htbp]
\vskip 4mm
\begin{center}
\includegraphics[width=8cm]{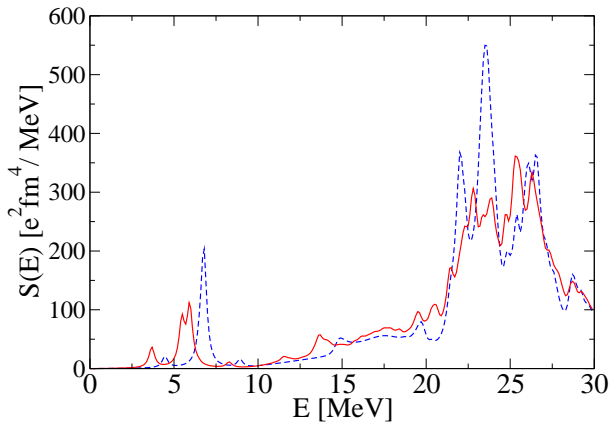}
\end{center}
\caption{\label{fig:9} Isovector quadrupole strength in $^{132}$Sn. 
The dashed line represents the continuum RPA and the full line the QTBA result.}
\end{figure}

The compression of the single-particle spectrum has opposite consequences for isoscalar and isovector resonances within the QTBA \cite{Avd}. 
In the isoscalar case the ph-force has to be weaker compared to the conventional RPA and for the isovector case the force has to be stronger. 
In Fig. \ref{fig:3} the continuum RPA result for the E1 photoabsorption cross section in $^{132}$Sn (dashed) is compared with the QTBA result (full line) and the data (dashed-dotted). 
As shown in Table \ref{tab:t6} the energy calculated in RPA about 2 MeV below the experimental mean energy and the QTBA 2.5 MeV. 
If we increase the isovector interaction by 10\% (QTBA*) we obtain perfect agreement with the data \cite{GSI}. 
The same holds for the width. 
In Ref. \cite{Avd} a different configuration space and a different number of phonons was used. 
The exact continuum influences the QTBA results most strongly beyond 20 MeV. 
In Fig.\ref{fig:4} and Table \ref{tab:t7} the RPA and QTBA results for $^{208}$Pb are shown. 
As in the case of $^{132}$Sn the theoretical cross section agrees excellently with the data if we slightly increase the isovector the isovector force as shown in Fig. \ref{fig:5}.
Finally we also calculated the giant quadrupole resonances in $^{208}$Pb and $^{132}$Sn. 
The results for $^{208}$Pb are shown in Fig. \ref{fig:6} and Fig. \ref{fig:7} and in Table \ref{tab:t8} and that for $^{132}$Sn in Fig. \ref{fig:8} and Fig. \ref{fig:9}. 
Our theoretical results for $^{208}$Pb are in fair agreement with the experimental values.
The numbers denoted by QTBA* are calculated with a slightly (5\%) increased isoscalar ph-force. 
In this special case the theoretical values agree nearly quantitatively with the data. 

\section{Conclusion}
In the present paper we continue the self-consistent investigation of nuclear structure properties in the frame-work of a theory which is based on the RPA but considers in a consistent way the effects of phonons. 
We concentrated on the two double magic nuclei $^{132}$Sn and $^{208}$Pb. 
Our model includes in a consistent way 1p1h configurations and phonons. 
The major change compared to our previous paper \cite{Avd} concerns the exact treatment of the single-particle continuum. 
This not only influences the strength distributions in the continuum but also allows a better definition of the size of the configuration space.
For nuclei far from the stability line the exact treatment will be crucial for quantitative predictions. 
Results using a discretized continuum depend on the details of the discretization prescription.
We calculated not only low- and high-lying collective states, but investigated in addition the low-lying non-collective states which represent the major part of the spectrum. 
Here we demonstrated the crucial influence of the phonons by comparing the conventional RPA with the QTBA. 
We used the SLy4 parametrization of the Skyrme ansatz in our calculations, which has been adjusted to the ground state properties as well as to collective states obtained within the RPA.
SLy4 has an effective mass m$^*$/m = 0.7.
The inclusion of phonons leads to a compression of the too wide HF single-particle spectrum and improves in this way appreciably the agreement between theory and experiment.
In this respect our calculation confirms microscopically the investigations by Mahaux et al. \cite{MAHAUX} who discussed the effective mass in nuclei and the influence of the phonons on the single-particle spectrum. 
The present investigation shows that nuclear structure theories have to treat the excitation spectrum and the separation energies simultanously.
We also showed that with very small changes in the ph-interaction one can reproduce the data nearly  quantitatively.
From this we conclude that a small renormalization of the SLy4 parametrization may give an optimal force
for self-consistent nuclear structure calculations which include the phonon degrees of freedom.

\begin{acknowledgement}
One of us (JS) thanks Stanislaw Dro\.zd\.z for many discussions and the Foundation for Polish
Science for financial support through the \emph{Alexander von Humboldt Honorary Research Fellowship}.
The work was also partly supported by the DFG and RFBR grants Nos. GZ:432RUS113/806/0-1 and 05-02-04005 and by
the INTAS grant No.03-54-6545.
\end{acknowledgement}

\end{document}